\documentstyle[12pt]{article}
\newcommand{\pT}{p_{T}}

\newcommand{\Hgg}{H  \to \gamma\gamma}
\newcommand{\signal}{g g \to \gamma\gamma g}
\newcommand{\sigdec}{g g \to H \to \gamma\gamma}
\newcommand{\be}{\begin{equation}}
\newcommand{\ee}{\end{equation}}

\def\eg{\hbox{\it e.g.}{}}    
\relax

\def\figcap{\section*{Figure Captions\markboth
     {FIGURECAPTIONS}{FIGURECAPTIONS}}\list
     {Fig. \arabic{enumi}:\hfill}{\settowidth\labelwidth{Fig. 999:}
     \leftmargin\labelwidth
     \advance\leftmargin\labelsep\usecounter{enumi}}}
 \relax
\def\reflist{\section*{REFERENCES\markboth
     {REFLIST}{REFLIST}}\list
     {[\arabic{enumi}]\hfill}{\settowidth\labelwidth{[999]}
     \leftmargin\labelwidth
     \advance\leftmargin\labelsep\usecounter{enumi}}}
 \relax
\def\tabcap{\section*{Tables\markboth
     {TABLES}{TABLES}}\list
     {Table \arabic{enumi}:\hfill}{\settowidth\labelwidth{Table 999:}
     \leftmargin\labelwidth
     \advance\leftmargin\labelsep\usecounter{enumi}}}
 \relax

\begin{document}
\begin{titlepage}
 \null
 \vskip 0.5in
\begin{center}
 \makebox[\textwidth][r]{IP/BBSR/99-11}

 \vspace{.15in}
  {\Large
   Production of Two Photons and  a Jet Through   Gluon Fusion
    }
  \par
 \vskip 1.5em
 {\large
  \begin{tabular}[t]{c}
    Pankaj Agrawal  \\
\em Institute of Physics \\
\em Sachivalaya Marg \\
 \em Bhubaneswar, Orissa 751005 India\\
   \em and\\
 Glenn Ladinsky \\
\em Michigan State University \\
 \em East Lansing, Michigan 48824, USA \\
\vspace{0.2cm}
  \end{tabular}}
 \par \vskip 5.0em
 {\large\bf Abstract}
\end{center}
\quotation

    We have computed the cross-section and distributions for
  the production of two photons and a jet via the process
 $g g \to \gamma\gamma g$.  In the Standard Model, this process 
 occurs at the one-loop level through pentagon and box diagrams. We show that
 at the LHC this is a significant mechanism for the
 production of the two photons and a jet. This process is also an
 important background to the Higgs boson production, when the Higgs
 boson decays into two photons. We also explore 
this process at the Tevatron.    

\endquotation
\vspace{1.0in}
 \baselineskip 10truept plus 0.2truept minus 0.2truept

\vfill
\mbox{DECEMBER 1998}
\end{titlepage}

\baselineskip=21truept plus 0.2truept minus 0.2truept
\pagestyle{plain}
\pagenumbering{arabic}
  
   The standard model has been extremely successful in describing
  a wide variety of phenomena in quite diverse situations. However,
  as colliders cross new energy and luminosity frontiers, there
  will be opportunities to test the model in new domains. In such
  a domain there is not only a potential to observe 
  beyond-the-standard-model processes, but also to observe those standard
  model processes that were not properly accessible earlier. Comparison of the
  the features of these processes with the standard model predictions
  will provide another avenue to test the standard model. Such processes
  will typically be either multi-particle final state processes or those
  that do not occur at tree level.

  In this letter, we discuss the production of two photons
  in association with a jet.  Observation of such events 
  provides an avenue for testing the QCD. Two photon + jet
  events have already been produced at the Tevatron
  and are expected to be produced more copiously at the
  LHC.  To compare the data with theory, one must compute
  the contribution of the process that makes the largest -- or
  at least significant --  contribution to the total cross-section. 
 At the LHC, the dominant mechanism is
  expected to be $ \signal$ as we discuss below. 
  Even at low energy machines such as the Tevatron and its upgrades, 
 low $ \pT$ events are
 also expected to receive significant contribution from
 the gluon initiated processes. Because
  of the technical complexity, the contribution of this
  process has not been computed thus far. We report
  the results of our computation for this process.

 Two photon + jet events
  can also be a background to the signature of the Higgs
  boson in the Standard Model, or a similar particle 
  in the models that have extended Higgs sector, \eg,
  supersymmetric standard model. One of the signatures
  of the Higgs boson that will be explored results through
  the decay mode $ \Hgg$. The CMS and ATLAS collaborations
  at the LHC are designing their detectors so that they could
  identify the Higgs boson through this decay mode, if such
  a Higgs boson exists.  The signature that one looks for
  results from  $ \sigdec$. In analyzing the feasibility of this
  signature the strong interaction corrections to the signal are included;
  such corrections are quite significant and could increase
  the leading order cross-section by about a factor of $2$. However, in  
  the background estimates the process $ \signal$ is not
 included because of the unavailability of the results. Clearly,
 to properly assess the feasibility of the two photon signature,
 one must include the contribution of the process that we
 compute.

 At the leading order, the tree level processes, $q \bar{q} \to \gamma\gamma g, 
  q g \to \gamma\gamma q, \bar{q} g \to \gamma \gamma \bar{q}$ contributes
  to the two photon and a jet events. These processes have already been
  computed. The next to leading order QCD correction to these
  processes have also been computed. At the Tevatron, because of
  small gluon luminosity, the above processes are the major contributor
  to the two photon and a jet events. However at the LHC, with large
  gluon luminosity, the process $ \signal$, though a one-loop process,
  is expected to dominate. Such expectation has already borne out
  in the context of two photon events production at the LHC, where the
  mechanism $ g g \to \gamma\gamma$, a one-loop process, is comparable
  to the tree-level process $ q \bar{q} \to \gamma\gamma$. Even at the
  Tevatron, the small $p_T$ events may get measurable contribution
  from the gluon-fusion mechanism; this is because such event are
  small-$x$ events where gluon density may not be negligible.

   The computation for the process $ \signal$ becomes complicated
   because of the number of diagrams and the length of expressions
   for the contribution of the each diagram to the amplitude of the
   process. There are $42$ diagrams that contribute to the process 
  for each quark flavor in the loop. 
  We can broadly classify these diagrams into two categories: a) $ 24$ pentagon
  diagrams, and b) $ 18$ box diagrams.  A representative diagram from
  each category is displayed in Figs. $1$ and $2$ respectively.
  Other diagrams can be arrived at by appropriate permutations of 
  external particle lines. All these diagrams have a quark-loop.
  Since we are interested in large $ \pT$ events, we have neglected the
  quark masses. Only quark that is expected to have large mass is  the top
   quark. However, because of the decoupling theorem that holds for
  the QED and QCD [\ref{dec}], we would expect the diagrams with the top-quark
  to make very small contribution. Furthermore, a pentagon loop has
  five quark propagators; inclusion of the quark mass will increase 
  the length of the amplitude by manifold, thus increasing the run-time
  of the code by a factor of  10--100, without significantly changing
  our results. One consequence of neglecting the quark-mass
  is the appearance of mass-singularities in the expressions of the 
  pentagon and box integrals. However, we can use this to our
  advantage. Since our cross-section should be free of mass-singularities,
  a check for the cancellation of such singularities allows us to check
  various parts of our code.

    In addition to mass singularities, some diagrams have ultraviolate
  singularities. The pentagon diagrams are ultraviolate finite, but box diagrams 
  individually have ultraviolate divergences. However, as the process
  at this order is ultraviolate finite, so ultraviolate divergences
  cancel. The verification of this cancellation is a check on
  our calculations. We describe other checks below.

    The calculational procedure for the process is as follows.
    A pentagon diagram calculation requires to find the trace
    of ten gamma matrices and do associated Lorentz algebra;
    while a box diagram computation requires to find the trace
    of eight gamma matrices and do the algebra. We perform this
    part of calculation using the symbolic manipulation program
    Form [\ref{form}]. The contribution of these diagrams
    is written in terms of tensor integrals. For our process, 
    tensor integrals can be of pentagon, box, triangle or bubble type;
    these integrals have five, four, three or two propagators
    respectively.
    As an example, most complicated integral in this calculation is
    a five-tensor pentagon integral:

 \begin{eqnarray}
    \label{eq:1}
   {\rm Penta5} = \int { d^4 k \over (2 \pi)^4} {k^\mu k^\nu k^\rho
    k^\sigma k^\delta \over \sf{P_1 P_2 P_3 P_4 P_5} } 
 \end{eqnarray}

   where,

 \begin{eqnarray}
    \label{eq:2}
   \sf{P_1} & = & (k^2 - m^2), \\
   \sf{P_2} & = & ((k+p1)^2 - m^2),\\
   \sf{P_3} & = & ((k+p1+p2)^2 - m^2) ,\\
   \sf{P_4} & = & ((k+p1+p2+p3)^2 - m^2),\\
  \sf{P_5} & = & ((k+p1+p2+p3+p4)^2 - m^2). 
  \end{eqnarray}

 Here $p1, p2, p3, p4$ and $p5$ are external four-momenta and $k$ is
 the loop-momentum. $m$ is the mass of the particle in the loop.
 Similarly one can define tensor integrals of other categories.


       In the amplitude of our process, the mass $m$ will be
    the mass of the quark in the loop. At the energy scales
   under consideration, $p_T^{min} \sim 20$ GeV, we can neglect
   the masses of the up, down, strange, charm and bottom
   quarks. As mentioned above, on the basis of decoupling 
   theorem, we can ignore
    the diagrams where top quark is in the loop. 
    These various integrals have ultraviolet singularities,
   as well as mass singularities. We have used dimensional
   regularization for the ultraviolet divergence. Mass singularities 
   have been regularized by using a small quark mass.

     To compute these tensor integrals, we have used the Oldenborgh-Vermaseren
   techniques [\ref{ov}] to reduce the tensor integrals to scalar integrals. To
   this end we encoded their algorithms using the 
   Mathematica system [\ref{math}].
   Resulting expressions were converted into Fortran programs. 
   We have opted for these techniques over more conventional
   Passarino-Veltman techniques [\ref{pv}] because of the compactness and its
   use of Gram determinants which may lead to more numerically stable
   results because of potentially including large cancellations.
   In the case of tensor-Pentagon integrals, we can write these 
   integrals in terms of Box tensor-integrals and Levi-Civita tensors
   in very compact way. More details of this part of calculations
   can be found in the Ref. [\ref{penta2}].

     Since pentagon diagrams individually are ultraviolet finite, their
     contribution to the amplitude involves only those 
     tensor-integrals that are ultraviolet finite.These include 
     five-tensor to scalar pentagon-integrals and some of the box 
     and triangle tensor-integrals.However, when these
     tensor integrals are written in terms of scalar integrals,
     then some terms due to the bubble scalar integral have 
     ultraviolet divergent pieces. However, these divergent
     terms cancel, thus giving a result without such divergences.
     We have checked that such divergences indeed cancel.
     In addition to ultraviolet singularities, there are mass
     singularities in scalar integrals. We have verified that
     these singularities cancel at the amplitude level. In an
     appendix to this letter, we have given the expressions for
     a few scalar integrals, in the limit of the small mass. There
     we can see the singularity structure.

      Individual box diagram contribution to the amplitude has
     both ultraviolet and mass singularities.Therefore,
     even ultraviolet divergent tensor integrals, such as
     four-tensor box integral, contribute to the amplitude.
     However, we have checked these both kind of singularities
     cancel in the amplitude. Apart from checking that the amplitude
     is singularity free, we have verified that the amplitude
     is  gauge invariant.
     For this purpose, we replace the polarization vector of a
     photon with that of its four-momentum. And find that the
     resulting expression vanishes.

     As we have discussed earlier, $42$ Feynman diagrams contribute
   to the amplitude. Given the number of the diagrams and the number
   of terms in the contribution of each diagram, it is not possible
   to analytically square the amplitude. We therefore evaluate the
   amplitude numerically, before taking its absolute value square.
   We have carried out numerical calculations for the Tevatron and the
   LHC.In each case, we compute the cross-sections and the distributions.
   One of the problem in the computation has been the slowness of the
   code. This happens because of the number of diagrams, sum over
   polarization and the lengthy expressions for the amplitudes and
   pentagon tensor-integrals. To solve this problem, we used a
   parallel version of the VEGAS algorithm [\ref{veseli}] that uses
   the PVM software system [\ref{pvm}].

We shall discuss the results at the LHC and the Tevatron. The center of mass
energy for the LHC is taken to be $ 14$ TeV; for the Tevatron the
center of mass energy is $2$ TeV. After computing the
parton level differential cross-section we convolute it with the CTEQ4
parton distributions [\ref{cteq4}]. We have chosen the leading order fit, set 3,
of the distributions. This is because, though the process is one-loop 
process but it is at leading order. These distributions are evolved
to the scale $Q = p_{T}^{min}$. We have applied the following generic
cuts: 

\begin{eqnarray}
  \label{eq:4}
& &  p_{T}^{\gamma} > p_T^{min}; \;\;\;\;\;\;
  p_{T}^{j} > p_T^{min} ; \;\;\;\;\;\;
  |\eta|^{\gamma, j} < 2.5; \nonumber \\
 & & \;\;\;\;\;\;\;\;\;\;\;  \Delta R(\gamma, \gamma) > 0.6;
   \;\;\;\;\;\;  \Delta R(\gamma, j) > 0.6;
\end{eqnarray}

   where $\eta$ is pseudo-rapidity and
 $\Delta R = \sqrt{(\Delta \eta)^2 + (\Delta \phi)^2 }$.

Some Results of our calculation 
are displayed in Figs. $3$ and $4$. Here we have given the cross-sections
as a function of the $p_T^{min}$. We clearly see that with
the expected yearly integrated luminosity of about $2$ fb$^{-1}$
at the Tevatron and about $10$ fb$^{-1}$ at the LHC, we could
have few hundred to few thousand events at the coming colliders.
We also note that with the increase in $p_T^{min}$ value, the
cross-section decreases steeply. This is as one would expect.
The major sources of uncertainties in our calculations are values of
gluon distributions and the scale at which these are evaluated.
These overall uncertainties are expected to be of the order of $20-30\%$.
We estimate this by varying the Q and also using leading order MRST
parton distributions [\ref{mrst}]. Results in more detail will be
presented in a longer write-up [\ref{penta2}].

  In conclusion, we have computed the cross-section and some
  distributions for the process $\signal$. We find that at the LHC
  as well at the Tevatron this process will give rise to measurable 
  number of events.
   Apart from a test of the QCD, these results have a bearing on the
  Higgs boson searches by a `two photons' signature. The background
  due to the process $\signal$ should be properly included to search
  the Higgs boson through `two photons' signature.

\begin{center}
{\bf  Appendix} 
\end{center}

  In the notations of the Ref [\ref{pv}] we have following expressions
 for a few scalar integrals.
  
  In the limit when $m^2 < s $, a triangle scalar integral 
  that we use is:

  \begin{eqnarray}
    \label{eq:7}
       C_0(m^2,m^2,m^2,0,0,s) = {i \pi^2 \over 2 s} [ ln^2(m^2) - 
           2 ln(m^2)  ln(-s -i \epsilon) + ln^2(-s -i \epsilon) ]
  \end{eqnarray}

 When $ s \ll m^2 $, then a bubble scalar integral is:

 \begin{eqnarray}
   \label{eq:13}
    B_{0}(m^2, m^2, s) = i \pi^2 [-{2 \over \varepsilon} - \gamma_E
                           + ln(4 \pi \mu^2) + 2 - ln( -s -i \epsilon)]
 \end{eqnarray}

  Only bubble-scalar integrals have ultraviolate singularity. 
 Triangle, box and pentagon scalar integrals have mass singularities.
 Here $\varepsilon = n -4$. As mentioned in the text, we have used
 dimensional regularization for the ultraviolate singularities and
 a small mass to regulate the mass singularities.

\relax
\def\pl#1#2#3{
      Phys.~Lett.~B {\bf #1},  #2 (#3)}
\def\zp#1#2#3{
      Z.~Phys.~C {\bf #1}, #2 (#3)}
\def\prl#1#2#3{
      Phys.~Rev.~Lett. {\bf #1}, #2 (#3)}
\def\rmp#1#2#3{
      Rev.~Mod.~Phys. {\bf #1}, #2 (#3)}
\def\prep#1#2#3{
      Phys.~Rep. {\bf #1}, #2 (#3)}
\def\pr#1#2#3{
      Phys.~Rev.~D {\bf #1}, #2 (#3)}
\def\epj#1#2#3{
      Eur.~Phys.~J.~C {\bf #1}, #2 (#3)}
\def\np#1#2#3{
      Nucl.~Phys.~B {\bf #1}, #2 (#3)}
\def\ib#1#2#3{
      ibid. {\bf #1}, #2 (#3)}
\def\nat#1#2#3{
     Nature {\bf #1}, #2 (#3)}
\def\ap#1#2#3{
      Ann.~Phys. {\bf #1}, #2 (#3)}
\def\sj#1#2#3{
      Sov.~J.~Nucl.~Phys. {\bf #1}, #2 (#3)}
\def\ar#1#2#3{
     Ann.~Rev.~Nucl.~Part.~Sci. {\bf #1}, #2 (#3)}
\def\ijmp#1#2#3{
     Int.~J.~Mod.~Phys. {\bf #1}, #2 (#3)}
\def\cpc#1#2#3{
      Computer Physics Commun.  {\bf #1}, #2 (#3)}
\begin{reflist}

\item\label{dec} T. Applequist and J. carrazone, \pr{11}{2856}{1974}.

\item \label{form}J. A. M. Vermaseren, Form (1989).

\item \label{ov} G. J. van Oldenborgh and J. A. M. Vermaseren, 
                \zp{46}{425}{1990}.

\item \label{math} S. Wolfram, Mathematica, Addison-Wesley Publishing
                  Company, (1996).

\item\label{pv} G. Passarino and M. Veltman, \np{160}{151}{1979}.

\item \label{penta2} P. Agrawal and G. Ladinsky, in preparation.

\item\label{veseli} S. Veseli, preprint FERMILAB-PUB-97/271-T.

\item\label{pvm} Message Passing Interface Forum, 
    Int.~J.~Supercomp.~Apps.
       {\bf 8}, 157 (1994) 

\item \label{cteq4} H. L. Lai {\it et. al.}, \pr{55}{1280}{1997}.

\item \label{mrst} A.D. Martin, R.G. Roberts, W.J. Stirling and R.S Thorne,
           \epj{14}{463}{1999}.

\end{reflist}

\newpage

\begin{figcap}

\item A pentagon type diagram for the process $ g g \to \gamma \gamma g$. 

\item A box type diagram for the process $ g g \to \gamma \gamma g$. 

\item Dependence of the cross-section for $p p \to \gamma \gamma  +$
          gluon  jet at the LHC on $p_T^{min}$. 

\item Dependence of the cross-section for $p \bar{p} \to \gamma \gamma  +$
    gluon  jet at the Tevatron on $p_T^{min}$.

\end{figcap}

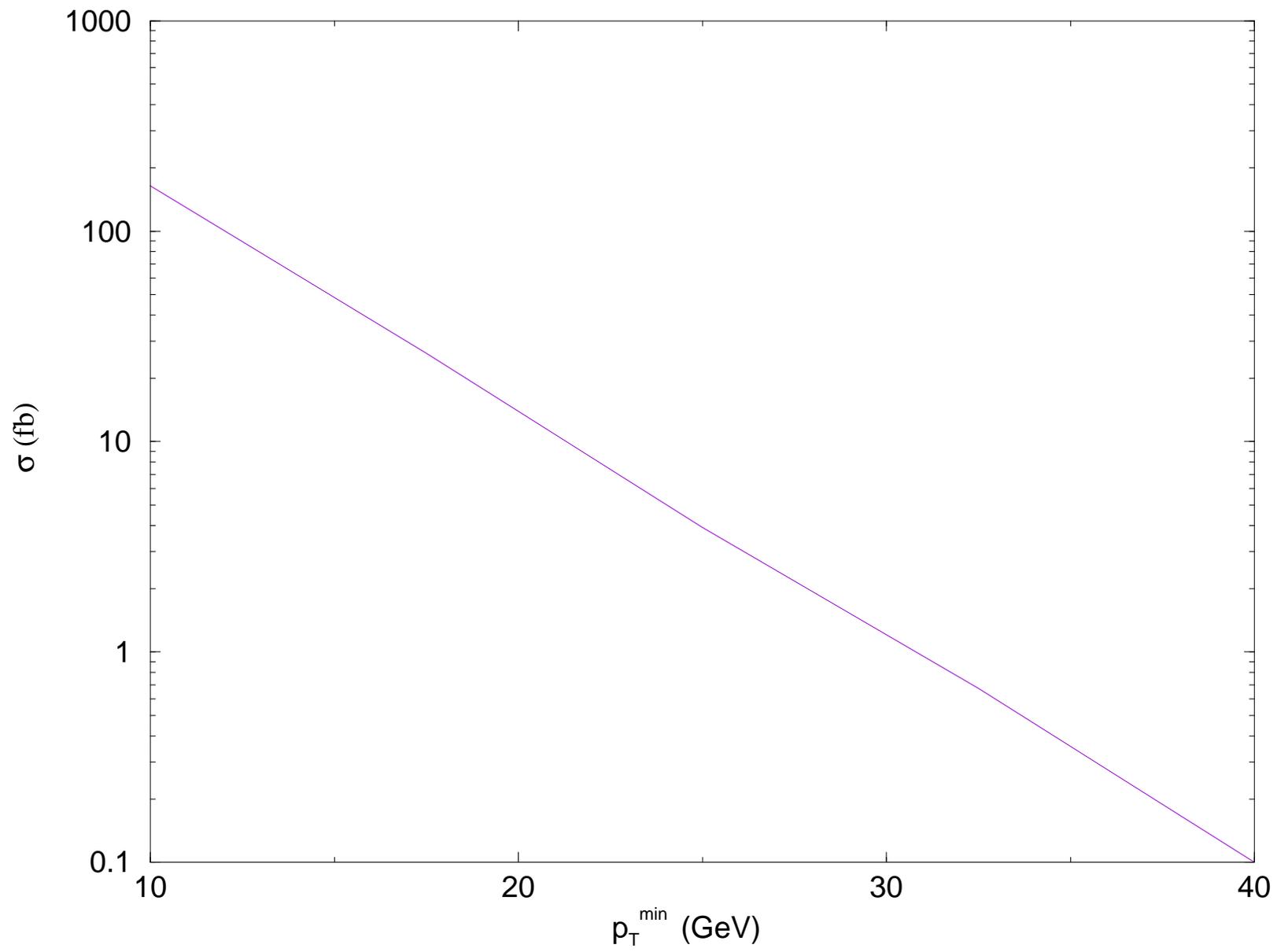
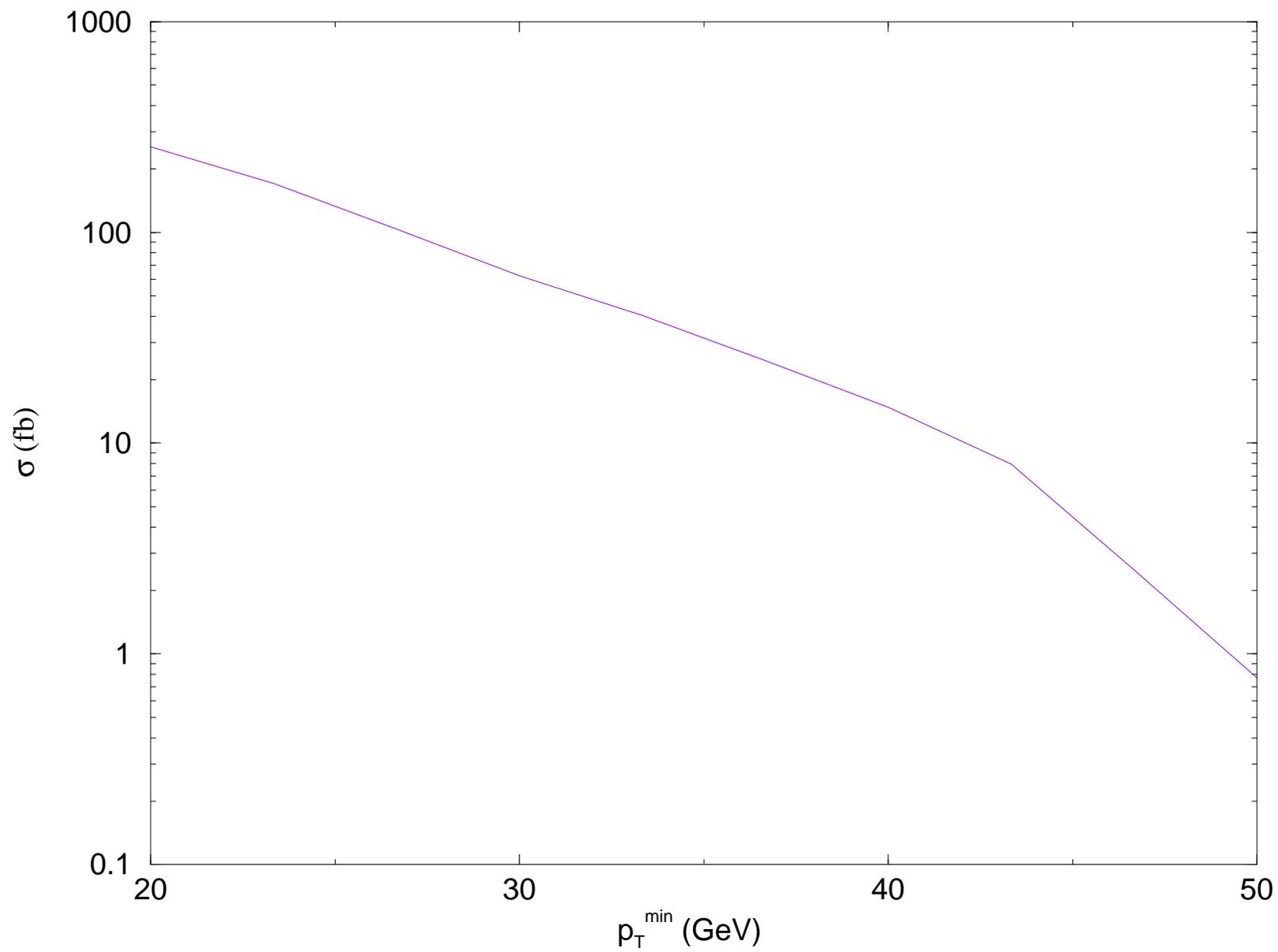
\end{document}